\documentclass[aps,twocolumn,noshowpacs, showkeys, superscriptaddress, floatfix, notitlepage]{revtex4-1}

\usepackage{graphicx}
\usepackage{dcolumn}
\usepackage{bm}

\usepackage{quantikz}
\usepackage{amsmath,amssymb} 
\usepackage{amsthm,epsfig,tikz}
\usepackage[justification=justified]{caption}
\usepackage{xcolor}
\usepackage{verbatim}
\usepackage[normalem]{ulem}
\usepackage[rightcaption]{sidecap}
\usepackage{subfigure}
\usepackage{forest,adjustbox}
\usepackage{multirow}
\usepackage{hyperref}  
\hypersetup{colorlinks,citecolor=black,urlcolor=black,linkcolor =black,hypertexnames=true,pdfproducer={Me}}
\usepackage{tikz}

\usepackage{float}
\makeatletter
\let\newfloat\newfloat@ltx
\makeatother
\usepackage{algorithm}
\usepackage{algcompatible}

\def\bra#1{\mathinner{\langle{#1}|}}
\def\ket#1{\mathinner{|{#1}\rangle}}

\renewcommand{\part}[2]{\frac{\partial #1}{\partial #2}}

\begin{document}

\title{Understanding the Mapping of Encode Data Through An Implementation of Quantum Topological Analysis}

\author{Andrew Vlasic}
\email{avlasic@deloitte.com}
\affiliation{Deloitte Consulting, LLP, Chicago, IL, 60606}

\author{Anh Pham}
\email{anhdpham@deloitte.com}
\affiliation{Deloitte Consulting, LLP, Atlanta, GA, 30303}

\begin{abstract}

A potential advantage of quantum machine learning stems from the ability of encoding classical data into high dimensional complex Hilbert space using quantum circuits. Recent studies exhibit that not all encoding methods are the same when representing classical data since certain parameterized circuit structures are more expressive than the others. In this study, we show the difference in encoding techniques can be visualized by investigating the topology of the data embedded in complex Hilbert space. The technique for visualization is a hybrid quantum based topological analysis which uses simple diagonalization of the boundary operators to compute the persistent Betti numbers and the persistent homology graph. To augment the computation of Betti numbers within a NISQ framework, we suggest a simple hybrid algorithm. Through a illuminating example of a synthetic data set and the methods of angle encoding, amplitude encoding, and IQP encoding, we reveal topological differences with the encoding methods, as well as the original data. Consequently, our results suggest the encoding method needs to be considered carefully within different quantum machine learning models since it can strongly affect downstream analysis like clustering or classification. 

\end{abstract}

\date{\today}

\maketitle

\section{Introduction}\label{sec:intro}

The power of quantum machine learning originates from the fact that classical data can be embedded in high-dimensional complex Hilbert space through the use of quantum circuits. Several strategies have been explored to encode classical data using different parameterized quantum circuits with different level of expressitivity to approximate the function representing classical data. For instance, Schuld et al. \cite{schuld2021effect} has shown that using Fourier series analysis can reveal if a variational quantum circuit is expressive enough to represent different classical data structure. However, a question still remains if encoding classical data using quantum circuits preserves the original structure of the data or it will change this structure after embedding in complex Hilbert space. 

As the applications of classical statistical modeling and machine learning modeling increase in popularity there has been a growing need of advanced exploratory data analysis to give insight to the stability of the model through the various structures of the data. Topological data analysis (TDA) has become a stable of general techniques to understand the geometric structure of data and assist with these insights. One main technique in TDA is that of Betti numbers, which give the number of ``holes" in each dimension. Classically, TDA has demonstrated to be a useful data analytic technique in revealing hidden geometrical structures of complex datasets through the analysis of Betti numbers and persistent homology \cite{wasserman2018topological}. 

The mathematical structure of calculating Betti numbers make this method a natural candidate for a quantum analog \cite{lloyd2016quantum}, denoted as qTDA. 
While there have been many advancements of the original quantum algorithm \cite{gyurik2020towards, siopsis2018quantum, hayakawa2021quantum, huang2018demonstration, gyurik2022towards}, the majority of these advancements assumes a universal quantum processor or a hybrid method with creative circuits. Taking into consideration the shortcomings of the NISQ era, this manuscript describes a hybrid qTDA method that is more scalable to augment the classical algorithm used to calculate Betti number. 

Motivated by the power of using topology to analyze the hidden geometrical structure in real world data, we addressed the question of how different quantum encoding techniques can affect the topology of embedded data structure within complex Hilbert space. Specifically, we explored the variation of persistent homology to understand the topological changes within different encoding techniques and with the original data. The empirical analysis displays that there are subtleties for each of the encoding techniques. In fact, our results reveal significant geometric differences between the original data and the encoding techniques, as well as geometric differences between each of the encoding techniques. Subsequently, this study implies further theoretical understanding is needed to derive the correct unitary operator to encode classical data beyond heuristic when designing a quantum variational circuit for machine learning application 

\section{Applications of Quantum to Real-World Data}\label{sec:app}

As quantum computing has become increasingly relevant as an advantage over classical computing there has been an increase in quantum analogs of machine learning algorithms. A few examples include linear regression \cite{lloyd2010quantum}, clustering \cite{basheer2020quantum, dang2018image}, utilizing kernels for support vector machines \cite{rebentrost2014quantum} with intuitive extensions to neural networks \cite{skolik2021layerwise, schuld2019quantum}, and generative modeling \cite{dallaire2018quantum}. As one may imagine there are situations were quantum is no better than classical methods \cite{lloyd2020quantum}. To understand how this may hold true consider modeling binary classification with a large feature space. If there is a clear delineation of the two classes within enough features a simple linear regression algorithm will model the data quite well. 

Given the nature of quantum algorithms many questions arise about how to incorporate classical data into a quantum circuit. A natural method to transfer continuous valued entries in a data record is to capture each value in the array as a rotation and capture each data point as an \textit{angle} \cite{schuld2019quantum, skolik2021layerwise}. Encoding the values of the data record to a quantum state through \textit{amplitude} is also quite natural. The authors in \cite{araujo2021divide} display a divide and conquer method, similar to that of the method in the seminal Grover-Rudolph paper \cite{grover2002creating}, which trades the number of qubits for the time to encode. This complexity is opposite of the more qubit less gate intensive method of angle encoding, or that of \textit{quantum Fourier transform} (QFT). The last method mentioned is \textit{IQP} \cite{havlivcek2019supervised} which is noted by the authors to leverage a quantum advantage, but the series of gates yields more of a neural network mapping than a true representation of the original data. There are other encoding techniques \cite{li2022quantum, wecker2015progress,sim2019expressibility}, for simplicity, these techniques are not considered.  

Mapping data to a quantum circuit has shown to be quite difficult and non-trivial \cite{schuld2019quantum}. Moreover, while these methods treat the data as a map to another Hilbert space the data is quite frankly mapped to a series of operators in the space of special unitary operators in dimension $2^n$, denoted as $\mathrm{SU}(2^n)$. However, there are explorations that are fairly in-depth in the analysis, which include authors in  \cite{huang2021power} that give criteria to when there is a quantum advantage in statistical modeling tasks, and the authors in \cite{schuld2021effect} explore a partial Fourier series of the operator encoded with a data point. To the authors' knowledge, the paper ``Effect of data encoding on the expressive power of variational quantum-machine-learning models" \cite{schuld2021effect} is the first effort to consider data from this perspective. This observation about data mapped to parameters of operators points to the question about what information, if any, is lost when data is encoded into a circuit.

\section{Encoding Approaches}\label{sec:encode}

The encoding methods noted in Section \ref{sec:app} will be discussed in detail to sample current methods and establish a deeper understanding. In a closed system there is a the periodic nature of quantum mechanics. From the periodic characteristics, it may be necessary to map the data point $D_i$ to a normalized form. There are many methods, such as dividing this vector by its magnitude, a significant amount relying on local information. Defining $\displaystyle \mathcal{S}^n = \left\{ x \in \mathbb{R}^{n} : x_i > 0 \ \forall \ i \ \mbox{and} \ \sum_{i=1}^{n-1} x_i = 1  \right\}$, another simplex, there exists a homeomorphic mapping of the form
\begin{equation}\label{eq:simp}
\begin{split}
    & f:\mathbb{R}^{n-1} \to \mathcal{S}^n , \\
     f(x) & = \frac{ \left( e^{x_1}, e^{x_2}, \ldots, e^{ x_{n-1} }, 1 \right)}{1+ \sum_{i=1}^{n-1} e^{x_i} }, \\
     f^{-1}(s) & = \big( \log(s_1/s_n ), \log(s_2/s_n ), \ldots, \log(s_{n-1}/s_n ) \big).
\end{split}
\end{equation}
 One may verify $f$ is a homeomorphism, and therefore, ensures no loss of information.

\textit{Angle} encoding is the first technique explored \cite{schuld2019quantum, skolik2021layerwise}. While this approach is intuitive has been noted to not fully leverage quantum, and in particular, each entry in the data point vector $D_i$ is mapped to a qubit, increasing the size of a typical register.  The encoded operator has the mapped form
\begin{equation}
     D_i \rightarrow \bigotimes_{l=1,2, \ldots,|D_i|} \exp( -i X_l D_i^l ) \ket{0\ldots0},
\end{equation}
where $X_{l}$ is the Pauli $X$ gate acting on the $l^{th}$ qubit and $D_i^l$ is the $l^{th}$ entry.

\begin{figure}
   \begin{adjustbox}{valign=t}
    \subfigure[ Binary Tree Decomposition]{ \begin{forest}
      for tree={l+=.25cm} 
      [$1$
        [$\sqrt{.55}$[$\sqrt{.2}$][ $\sqrt{.35}$ ]]
        [$\sqrt{.45}$[$\sqrt{.15}$][$\sqrt{.3}$]]
      ]
    \end{forest} }
    \end{adjustbox}
    \begin{adjustbox}{valign=t} 
    \subfigure[ Encoding of the Binary Tree]{ \begin{quantikz}[thin lines] 
        \lstick{$\ket{0}$} & \gate{R_y\Big(\sqrt{.45} \Big)}  & \octrl{1} &\ctrl{1}& \qw
       \\ \lstick{$\ket{0}$} & \qw & \gate{R_y\Big(\sqrt{.3/.55 }\Big)}&  \gate{R_y\Big(\sqrt{ .15/.45}\Big)} & \qw
    \end{quantikz} }
\end{adjustbox}
\caption{This is an example on how to amplitude encode data: (a) displays the binary tree decomposition of $( \sqrt{.2}, \sqrt{.35}, \sqrt{.15}, \sqrt{.3})$, and (b) is the respective circuit to encode the binary tree.}\label{fig:amp_enc}
\end{figure}
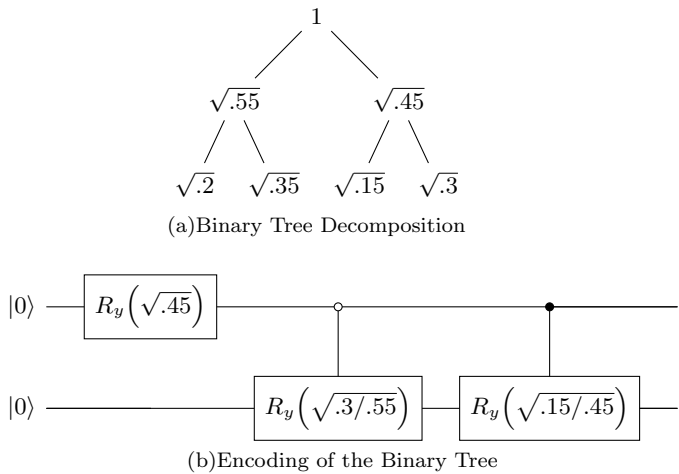

The \textit{amplitude} approach assumes all the data points $\tilde{D}_i \in \mathcal{S}^n$. As noted above, one may map any data point into this form without any loss of information. The authors in  \cite{araujo2021divide} display an efficient method to encode the amplitudes of the data, utilizing the method in the Grover-Rudolph \cite{grover2002creating} approach to load probability distributions. Given the intricate procedure of the technique it will not be discussed in full detail. Moreover, the authors give a detailed description of their algorithm, enabling one to implement this approach. For completeness, an example of how to implement this particular amplitude encoding is given. Take the simple example of encoding the conveniently prepared data point $( \sqrt{.2}, \sqrt{.35}, \sqrt{.15}, \sqrt{.3})$ which is decomposed into a binary tree by recursively splitting each piece of the data point in half with the goal of creating the state $\sqrt{.2} \ket{00} + \sqrt{.35} \ket{10} + \sqrt{.15} \ket{01} + \sqrt{.3} \ket{11}$. Each split is then normalized and the left node is taken as the parameter of the $R_y$ gate. This binary tree and subsequent circuit is shown in Figure \ref{fig:amp_enc}.

Lastly, the \textit{IQP} encoding approach \cite{havlivcek2019supervised} is discussed. The authors assume $x\in (0,2\pi]^n$ and note the approach suggests a quantum advantage. Taking the function $f$ in Equation \ref{eq:simp} and defining the function $\tilde{f} = 2\pi \cdot f$ that maps $\tilde{f}: \mathbb{R}^{n-1} \to (0,2\pi]^n$ does not lose information since $f$ is a homeomorphism. Denoting $Z_i$ as the Pauli $Z$ gate acting on the $i^{th}$ qubit the authors define the specific unitary operator \begin{equation}\label{used}
U_Z(x) = \exp\left( \sum_{i=1}^n x_i Z_i + \sum_{i=1}^n \sum_{j=1}^n (\pi - x_i)(\pi- x_j) Z_i Z_j \right)
\end{equation} for application, while giving the general operator as $\displaystyle U_{\Phi}(x) = \exp\left( \sum_{S \subset \{1,2,\ldots,n\}} \Phi_{S}(x) \prod_{i\in S}Z_i \right)$. One may note the coefficients in the quadratic terms are centered around $0$ with standard deviation of $1$, and the original data may be mapped accordingly into a different form. Denoting $H$ as the Hadamard gate, the IQP encoding is defined as 
\begin{equation}
D_i \rightarrow U_Z( D_i) H^{\otimes n} U_Z( D_i) H^{\otimes n} \ket{0\ldots0}.
\end{equation}
The authors derived this encoding by considering Ising interactions of the unitary operators in $U_Z$ and the Hadamard gates adding uniform superpositions.

\section{Topological Data Analysis and Quantum Computation}\label{sec:tda}

With the explosion of statistical modeling it has become more prevalent to understand the geometric structure for further insight into how to model the data, identify weak structures of underrepresented data to anticipate failure, and compare training data and data observed after training. Topological data analysis (TDA) has been exhibited to be a powerful tool to obtain this goal \cite{friedman1998computing,ghrist2008barcodes} but is quite computationally intensive. One particular method that is quite useful in classifying the geometric structure of the data is calculating the Betti number for various topological features. 

To understand what a Betti number represents more intuitively, consider a record of data points. First it is necessary to construct a \textit{simplicial complex}, which is a collection of the data points with an $\epsilon$-sized ball around each point that creates individual points, lines, triangles, tetrahedron, and corresponding iterative higher-level simplicial objects; $\epsilon$ is a hyperparameter and is intuitively called the \textit{grouping scale}. The collection of simplices created from parameter $\epsilon$, denoted as $S^{\epsilon}$, is known as a \textit{Vietoris-Rips simplicial complex}. After the construction of these objects the number of connected data points, one-dimensional ``circular holes", two-dimensional areas void of data points, and corresponding higher-dimensional voids. For $k\in \{0,1,2,\ldots\}$ the Betti number $b_k$ corresponds to the respective topological descriptions above. 

Betti numbers have a deeper mathematical description with homology, which is essential to describe and incorporate in a circuit. Given a data set $D_s$, denote $H_k^{\epsilon}(D_s)$ as the $k^{th}$ homology group of $D_s$ generated from the parameter $\epsilon$. The complete simplicial complex created from $\epsilon$ is defined $H^{\epsilon} = \bigoplus_k H^{\epsilon}_k(D_s)$. To connect individual simplices, define the \textit{boundary map} as $\delta_k:H_k^{\epsilon}(D_s) \to H_{k-1}^{\epsilon}(D_s)$, and given the derivation of the simplicial complex, one may see the natural mapping of $\delta_k$. Denoting the kernel of a function as $\mathrm{ker}$ and the image of a function as $\mathrm{Im}$, with this structure we may define the $k^{th}$ homology as the quotient space $H_k^{\epsilon}(D_s) = \mathrm{ker} \ \delta_{k}/\mathrm{Im} \ \delta_{k+1}$ and $b_k = \mathrm{dim} \big( H_k^{\epsilon}(D_s) \big)$.

This structure enables a derivation towards an generator of connectivity. Combinatorial Laplacians \cite{chung1996combinatorial} give the exact generator and has the form $\Delta_k= (\delta_{k})^{\dagger} \delta_{k} + \delta_{k+1} (\delta_{k+1})^{\dagger}$, and one may see that $\Delta_k$ is a Hermitian matrix. The combinatorial Laplacian may calculate the $k^{th}$ Betti number, $\beta_k$, by deriving
\begin{equation}\label{eq:betti_num}
    \beta_k = \mbox{dim} \ \mbox{ker} \ \Delta_k.
\end{equation} See \cite{carlsson2021topological} for further more in-depth information about Betti numbers. 

Since the boundary maps are linear this gives makes this algorithm a candidate for a quantum analog \cite{lloyd2016quantum}, denoted as \textit{qTDA}. In the quantum setting for the space $H_k^{\epsilon}$ is spanned by $\ket{s_k}$ where $s_k \in S^{\epsilon}_k$, where $S^{\epsilon}_k$ is the set of $k$-simplices. The boundary map applied to $\ket{s_k}$ has the form
\begin{equation}\label{eq:boundmap}
    \delta_k \ket{s_k} = \sum_{j} (-1)^j \ket{s_{k-1}(j)}
\end{equation}
where $s_{k-1}(j)$ is the $k-1$ simplex on the boundary of $s_k$ with the $j^{th}$ vertex removed from $s_k$.

Since the derivation of qTDA there have been many extensions \cite{gyurik2020towards,siopsis2018quantum,hayakawa2021quantum}. However, the general flow of the algorithm has been consistent; please see \cite{khandelwal2023quantum} for a brief overview. The outline of the algorithm is displayed in Algorithm \ref{alg:gen}. 

\begin{algorithm}
\caption{General qTDA}
\begin{algorithmic}[1]
\STATE $i \gets 1$
\WHILE{$i\leq L$}
    \STATE $\displaystyle \frac{1}{ \sqrt{|S_k^{\epsilon}|} } \sum_{ s_k \in S_k^{\epsilon}} \ket{s_k} \gets$ Grover's algorithm 
    \STATE $\displaystyle \frac{1}{ \sqrt{|S_k^{\epsilon}|} } \sum_{ s_k \in S_k^{\epsilon}} \ket{s_k}\otimes \ket{s_k} \gets$ copy states to eigenvalue registry with CNOT operations
    \STATE $\displaystyle \frac{1}{ \sqrt{|S_k^{\epsilon}|} } \sum_{ s_k \in S_k^{\epsilon}} \ket{s_k}\bra{s_k} \gets$ trace out the ancillary register 
    \STATE $\displaystyle e^{i \Delta^{D_s}_k } \gets$ apply unitary to eigenvalue registry
    \STATE Apply phase estimation to eigenvalue registry 
    \STATE Measure the eigenvalue register to readout the approximated eigenvalue $\Tilde{\lambda}$
    \STATE $i \gets i+1$
\ENDWHILE
\STATE \textbf{return} $|\{\Tilde{\lambda}: \Tilde{\lambda} = 0\}|/L$
\end{algorithmic}
\label{alg:gen}
\end{algorithm}

The general flow of the circuit side-steps how to incorporate real-world data into the circuit. Currently, there are two main methods: (1) calculate the distances (or pseudo-distances), apply the $\epsilon$ hyperparameter filter, make respective connections in classical computation then incorporate this final matrix into the circuit; or (2) encode the data into the circuit and do all calculations within the circuit. The first method is oriented for the NISQ-era, and in fact, may be faster given a small enough data with the available gates. Furthermore, the first method is difficult to incorporate if one would like to calculate the Betti number of the encoded data. The second method is designed for a universal QPU as the number of gates needed and sensitivity to the noise of qubits plays a huge factor in exact calculations.  

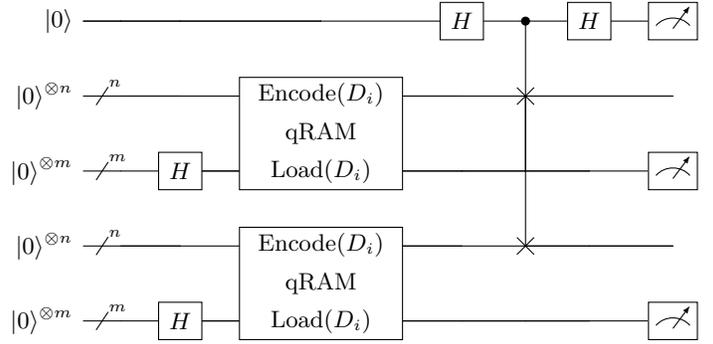
\begin{figure}[th]
\noindent
    \begin{quantikz}[thin lines] 
        \lstick{$\ket{0}$}& \qw & \qw &\qw & \gate{H}  & \ctrl{2} & \gate{H} &\meter{}
       \\ \lstick{$\ket{0}^{\otimes n}$}& \qwbundle[]{n} & \qw &  \gate[wires=2,disable auto height]{ {\begin{array}{c} \text{Encode}(D_i) \vspace{4pt} \\ \text{qRAM} \vspace{4pt} \\ \text{Load}(D_i) \end{array}} }& \qw & \swap{2} &\qw &\qw 
       \\ \lstick{$\ket{0}^{\otimes m}$}& \qwbundle[]{m}  & \gate{H} & \qw &\qw & \qw& \qw &\meter{}
       \\ \lstick{$\ket{0}^{\otimes n}$}& \qwbundle[]{n} & \qw &  \gate[wires=2,disable auto height]{ {\begin{array}{c} \text{Encode}(D_i) \vspace{4pt} \\ \text{qRAM} \vspace{4pt} \\ \text{Load}(D_i) \end{array}} } & \qw & \swap{-1} &\qw &\qw
       \\ \lstick{$\ket{0}^{\otimes m}$}& \qwbundle[]{m} & \gate{H} & \qw &\qw   & \qw & \qw  &\meter{}
    \end{quantikz}
\caption{This circuit displays a sub-process to calculate the inner product of two randomly chosen data points from the same uniformly distributed data points encoded into a quantum circuit and apply the \textit{SWAP test}.}
\label{fig:qram}
\end{figure}

While there are hardware shortcomings of the second method it is quite interesting to explore creating such a non-hybrid circuit. The rest of this section explores how one may implement comparing two randomly selected data points from a record. The method in \cite{liu2021rigorous} displays how to get the encoded inner product of two data records. Since $\braket{\psi}{\psi} = 1$ for all non-zero vectors in a circuit one may see that 
\begin{equation}\label{eq:inner-L2}
 \sqrt{ \sum_{k=1}^{n} |\psi^i_k - \psi^j_k|^2 } = \sqrt{2\cdot (1-| \braket{\psi^i}{\psi^j}|) }. 
\end{equation}
The authors in \cite{araujo2021divide} give a viable technique to implement a qRAM for the entire record of data. Coupling the qRAM and inner product with the technique in \cite{basheer2020quantum}, which utilizes the Hadamard test yields, a sub-process to compare randomly drawn data points with a quantum advantage. See Figure \ref{fig:qram} for an overview of such a circuit. 

\section{Hybrid Quantum Topological Data Analysis}\label{sec:hybrid}

While the NISQ-era inhibits a purely quantum solution to calculate Betti numbers, there have been efforts to create a hybrid solution \cite{huang2018demonstration}. In particular, the Huang et al. display a hybrid method, though the circuit given is a toy example, that calculates Betti number $b_1$ with five data points with the $L^2$-distance between each point. The authors then derive a creative circuit to calculate the $b_1$ score of the network of the boundaries. Such a circuit with pre-calculated boundaries has been previously noted \cite{lloyd2016quantum, siopsis2018quantum}. However, the authors go a step further and derive a matrix of the Equation \ref{eq:boundmap}. One may observe that with this matrix all is necessary to finish the Betti number calculation is to derive the eigenvectors and determine $|\{\Tilde{\lambda}: \Tilde{\lambda} = 0\}|$. 

\begin{figure}[th]
\noindent
    \begin{quantikz}[thin lines] 
       \lstick{$\ket{0}^{\otimes m}$} & \qwbundle[]{m} &  \gate{ \text{Encode}(D_i)} & \gate{ \text{Encode}^{\dagger}(D_j) } &\meter{}
    \end{quantikz}
\caption{A kernel method to calculate the absolute value squared of the inner product of two encoded data points, $D_i$ and $D_j$, where the data point $D_j$ is encoded with the inverse of the technique. Finally, measured in the computational basis. This is known as the \textit{fidelity test}.}
\label{fig:innrprd}
\end{figure}
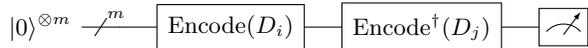

The method in \cite{huang2018demonstration} can be expanded by including data encoded into a circuit. However, one must be able to calculate the distance between each point in the record. A technique one may utilize is displayed in the circuit in Figure \ref{fig:innrprd}, which is known as the \textit{SWAP test}. This circuit yields the calculation $|\braket{E(D_j)}{E(D_i)}|^2$, where $E$ denotes the encoding method and is shortened notation for simplicity. One may derive this kernel method explicitly by simplifying the measurement of the circuit on the initial state $\ket{0\ldots0}\bra{0\ldots0}$ ,
\begin{equation}\label{eq:fidelity}
\begin{split}
     & \bra{0\ldots0}E(D_j)E(D_i)^{\dagger} \mathcal{M} E(D_j)^{\dagger} E(D_i)\ket{0\ldots0} \\
        & = \bra{0\ldots0} E(D_j)E(D_i)^{\dagger} \ket{0\ldots0} \\
        &\times \bra{0\ldots0} E(D_j)^{\dagger} E(D_i) \ket{0\ldots0}  \\
        & = |\bra{0\ldots0} E(D_j)^{\dagger} E(D_i) \ket{0\ldots0}|^2 \\
        & = |\braket{E(D_j)}{E(D_i)}|^2.
\end{split}
\end{equation}
A few kernel methods with calculation are described in the Pennylane \cite{bergholm2018pennylane} documentation.

\begin{algorithm}
\caption{Hybrid qTDA}
\raggedright\textbf{Input:}  Betti number $k$
\begin{algorithmic}[1]
\STATE $B \gets$ calculate inner products matrix 
\STATE $B^{\epsilon} \gets B$ apply $\epsilon$ filter 
\STATE $\{\partial_k,\partial_{k+1} \}  \gets B^{\epsilon}$ calculate the boundary operators 
\STATE $\Delta_k \gets \partial_k^{\dagger}\partial_k + \partial_{k+1}\partial_{k+1}^{\dagger}$ calculate combinatorial Laplacian
\STATE \textbf{Either} Decompose $\Delta_k$ as hamiltonian composed of Pauli matrices using Pauli decomposition, and feed that into the VQD sub-circuit and measure $L$ times to readout the approximated eigenvalues $\Tilde{\lambda}$ then \\ \textbf{return} $|\{\Tilde{\lambda}: \Tilde{\lambda} = 0\}|/L$ 
\STATE \textbf{Or} feed $\Delta_k$ into classical eigensolver then \\ \textbf{return} $|\{\Tilde{\lambda}: \Tilde{\lambda} = 0\}|$ 
\end{algorithmic}
\label{alg:hybrid}
\end{algorithm}

The proposed hybrid algorithm is given in Algorithm \ref{alg:hybrid}. While many of the steps are classical, Step 1 requires a quantum circuit for encoded data points, and the last step is to calculate the eigenvalues, recalling from Equation \ref{eq:betti_num} that $\Delta_k$ is Hermitian. Higgott, Wang, and Brierley in \cite{higgott2019variational} derive a circuit, noted as the \textit{Variational Quantum Deflation} (VQD), which is a NISQ-era friendly implementation to calculate the spectrum of a Hamiltonian. While there are other circuits that may be utilized the calculate the entire set of eigenvalues, given the proliferation of the algorithm, VQD will be noted as the preferred algorithm. However, VQD will also be used as a place holder for similar quantum algorithms. Instances when there isn't a quantum advantage, for example when the circuit is too long or the matrix is small enough, one may then apply a classical eigensolver. 

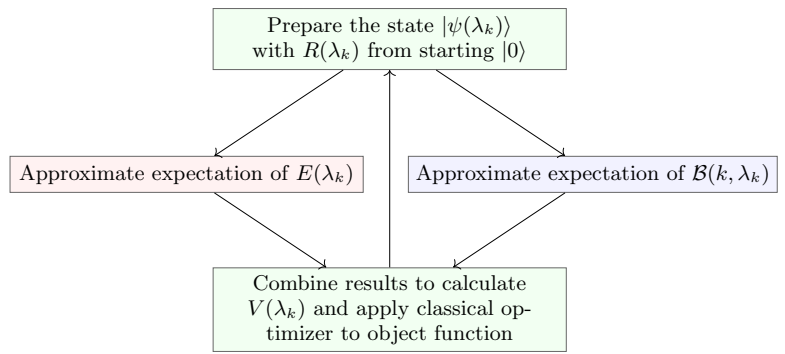
\begin{figure}[th]
\scalebox{.9}{
\begin{tikzpicture}[
squarednodeR/.style={rectangle, draw=black!60, fill=green!5, thin, text width=50mm,align=center},
squarednoder/.style={rectangle, draw=black!60, fill=red!5, thin, text width=50mm,align=center},
squarednodeb/.style={rectangle, draw=black!60, fill=blue!5, thin},
]
\node[squarednodeR] at (0,0) (main)  {Prepare the state $\ket{\psi(\lambda_k)}$ with $R(\lambda_k)$ from starting $\ket{0}$ };
\node[squarednoder] at (-3,-2) (ham)  {Approximate expectation of $E(\lambda_k)$};
\node[squarednodeb] at (3,-2)  (over) {Approximate expectation of $\mathcal{B}(k,\lambda_k)$};
\node[squarednodeR] at (0,-4) (class)  {Combine results to calculate $V(\lambda_k)$ and apply classical optimizer to object function };
\draw[->] (main) -- (ham);
\draw[->] (main) -- (over);
\draw[->] (ham) -- (class);
\draw[->] (over) -- (class);
\draw[->] (class) -- (main);
\end{tikzpicture}
}
\caption{Variational algorithm schematic of the VQD process.}\label{fig:vqd}
\end{figure}

For completeness, the VQD algorithm is explained. VQD was derived as an extension of the variational quantum eigensolver (VQE), see \cite{tilly2022variational} for an overview. Given a Hamiltonian $H = \sum c_j P_j$, VQE starts with a real $\lambda$ the ansatz state $\ket{\psi(\lambda)}$ and seeks to minimize the expectation $\displaystyle E(\lambda) := \sum c_j \bra{\psi(\lambda)} P_j  \ket{\psi(\lambda)}$, denoted as $\lambda_0$, approximating the ground state. To calculate the $k^{th}$ state Higgott, Wang, and Brierley in \cite{higgott2019variational} created the cost function
\begin{equation*}
\begin{split}
    V(\lambda_k) & := \bra{\psi(\lambda_k)} H \ket{\psi(\lambda_k)} + \sum_{i=0}^{k-1} \beta_i |\braket{\psi(\lambda_k)}{\psi(\lambda_i)} |^2 \\
    & := E(\lambda_k) + \mathcal{B}(k,\lambda_k) 
\end{split}
\end{equation*}
where the $\beta_i$ values are chosen sufficiently large to ensure orthogonality, $|\braket{\psi(\lambda_i)}{\psi(\lambda_j)} |^2 = 0$ for $i \neq j$. Denote $R(\lambda_k)$ as the procedure to prepare the circuit. The algorithm starts with an initial guess then iterates until a designated decision to stop. The schematic is given in Figure \ref{fig:vqd} and is given in generality to adjust for evolving techniques. For instance, there are a number of ways in which to calculate the expectation including the fidelity test described in Equation \ref{eq:fidelity} or the ``destructive SWAP test" \cite{larose2019variational}.

\begin{algorithm}
\caption{VQD for Spectrum Approximation}
\raggedright\textbf{Input:} Calculate the first $K+1$ eigenvalues. 
\begin{algorithmic}[1]
\STATE Apply VQE to approximate $\lambda_0$
\STATE $i \gets 1$
\WHILE{$i\leq K$}
    \STATE $\lambda_i \gets$ apply procedure in Figure \ref{fig:vqd} 
    \STATE $i \gets i+1$
\ENDWHILE
\STATE  $\{\Tilde{\lambda} \} \gets$ approximate eigenvalue spectrum using $\lambda_0, \ldots, \lambda_k$ from $L$ measurements.  
\STATE \textbf{return}  $\{\Tilde{\lambda} \}$ 
\end{algorithmic}
\label{alg:vqd}
\end{algorithm}

\section{Empirical Analysis}\label{sec:emp}

\begin{figure}[ht]
    \centering
    \includegraphics[width=0.45\textwidth]{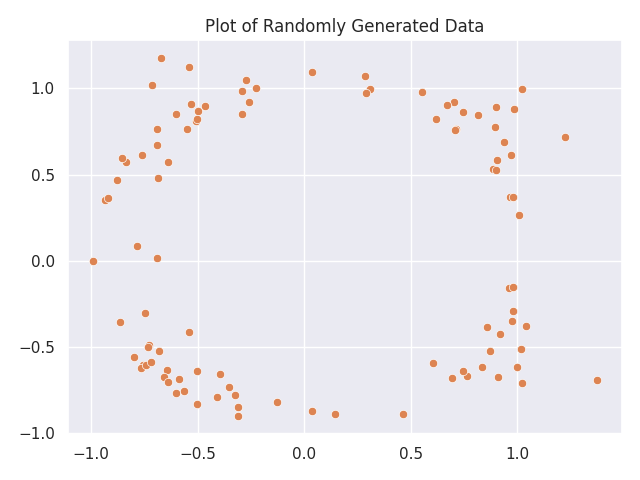}
    \caption{Scatter plot of the randomly generated two dimensional data analyzed.}
    \label{fig:scatter}
\end{figure}

\begin{figure}[ht]
    \centering
    \subfigure[]{\includegraphics[width=0.45\textwidth]{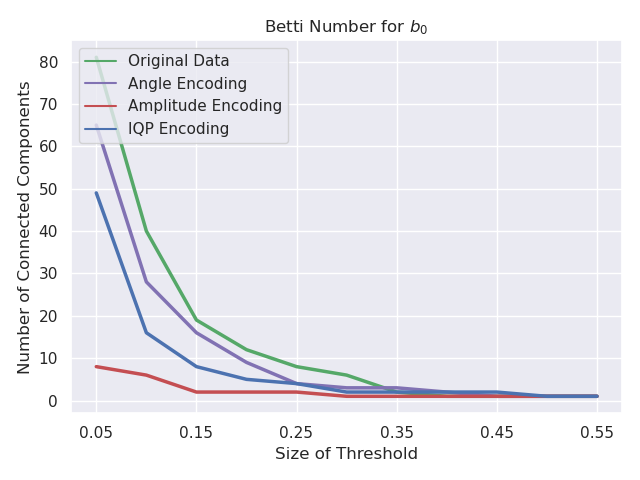}}
    \subfigure[]{\includegraphics[width=0.45\textwidth] {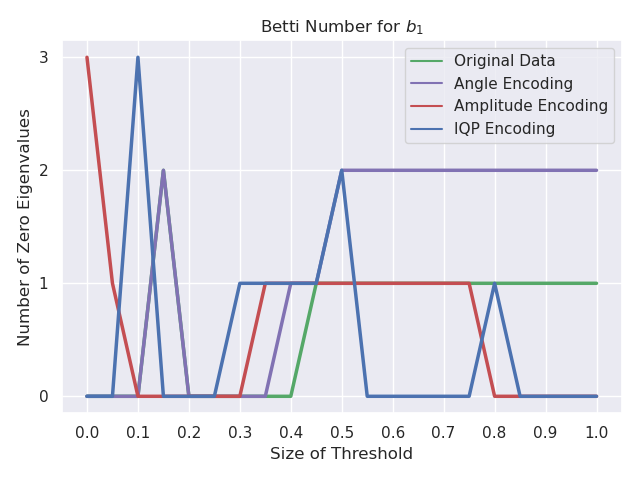}}
    \caption{ The Betti number for $b_0$ and $b_1$ are computed for the original data, angle encoded data, amplitude encoded data, and the IQP encoded data. Figure (a) displays Betti numbers for $b_0$ with the interval of $[.05,.55]$ with increments of $.05$, Figure (b) is the Betti numbers for $b_1$ with the interval of $[0.0,1.0]$ with increments of $.05$.}
    \label{fig:compare}
\end{figure}

To display the potential differences in the different encoding methods, one hundred randomly data points were generated and encoded with the angle method, amplitude method, and the IQP method.  The Euclidean distance is applied to the original data and the data encode with each of the three approaches. After the distances are calculated the Betti number for $b_1$ is derived at different threshold levels. While there are other methods, for simplicity these three methods are considered. Given the stark contrast shown in Figure \ref{fig:compare} between all of the data sources, it is believed other encoding methods will have a significant difference between the original data and other respective encoding techniques.

To calculate the Betti number for $b_1$ the algorithm described in Algorithm \ref{alg:hybrid} is utilized, however, given the size of the Hamiltonian matrices, the eigenvalues are calculated classically and the number of eigenvalues equal to $0$ are given. The Gudhi package \cite{gudhirips} was utilized to calculate the simplices and the respective faces.  

\begin{figure*}[th!]
    \centering
    \subfigure[]{ \includegraphics[width=0.4\textwidth] {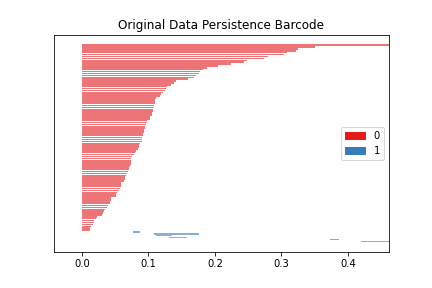} }
    \subfigure[]{ \includegraphics[width=0.351\textwidth] {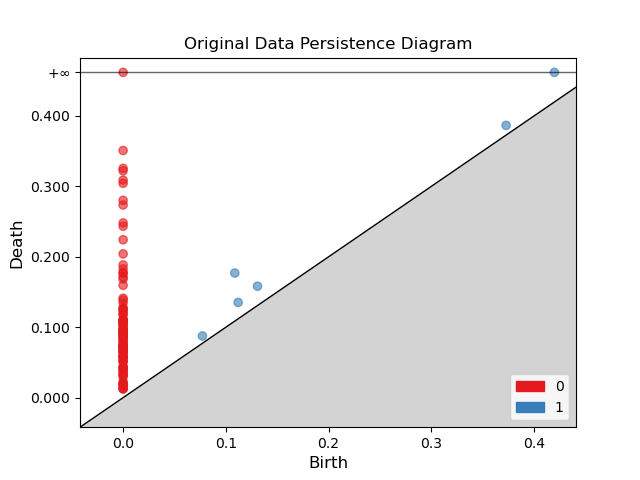} }
    \subfigure[]{\includegraphics[width=0.4\textwidth] {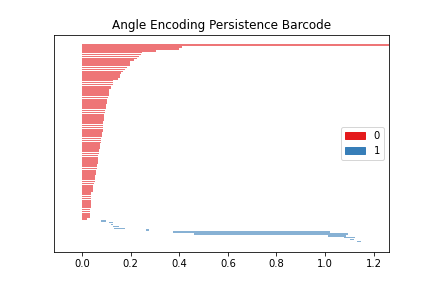}}
    \subfigure[]{\includegraphics[width=0.351\textwidth] {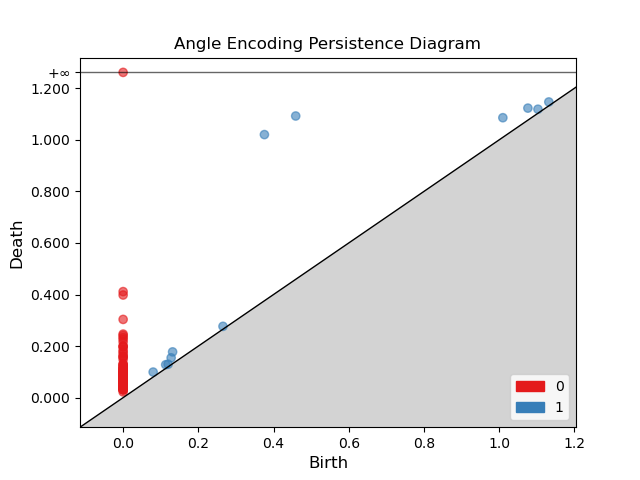}}
    \subfigure[]{\includegraphics[width=0.4\textwidth] {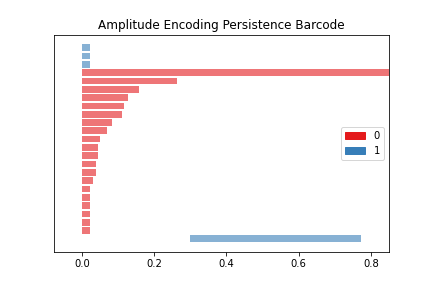}}
    \subfigure[]{\includegraphics[width=0.351\textwidth] {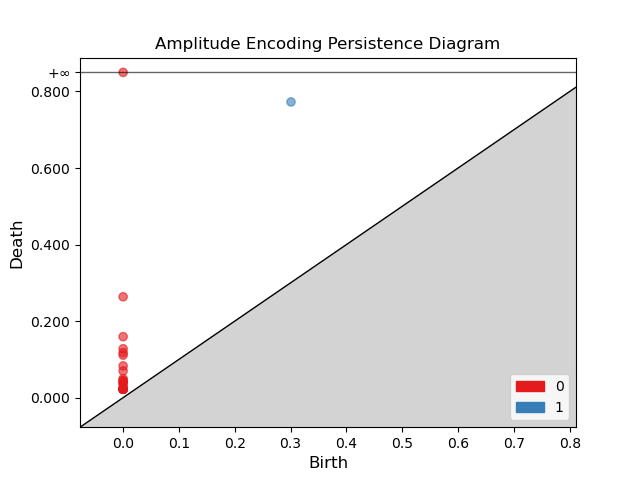}}
    \subfigure[]{\includegraphics[width=0.4\textwidth] {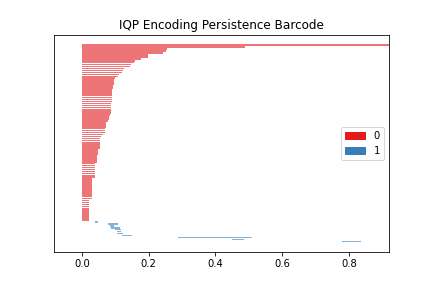}}
    \subfigure[]{\includegraphics[width=0.351\textwidth] {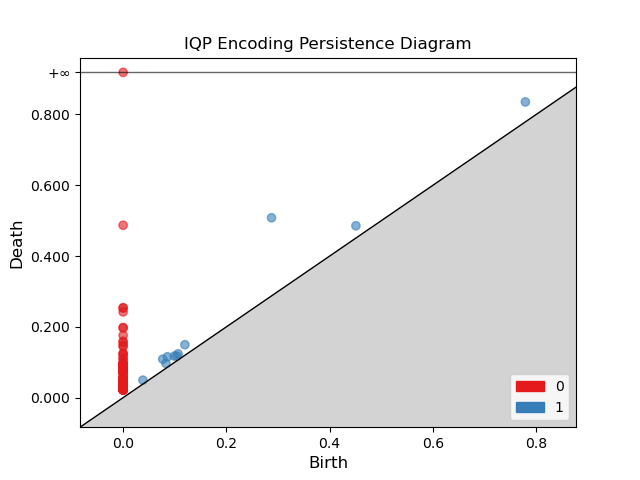}}
    \caption{The persistence barcode and respective diagram for each encoding method and original data. Each row gives insight into each respective geometry, demonstrating difference between all of the methods, as well as differences between each encoding method and the original data.}
    \label{fig:barcodes}
\end{figure*}

The data was generated with NumPy \cite{harris2020array}. One thousand two dimensional from a uniform distribution in the interval $[-1,1]$ was sampled with where each data point is normalized, one thousand two dimensional data points were then generated with a Pareto distribution with $\alpha = 10$, and finally these two sets are added together; see Figure \ref{fig:scatter} for a scatter plot of the data.

The circuits for each of the encoding methods were implemented with Qiskit \cite{aleksandrowicz2019qiskit} utilizing a simulator backend. For each pair of data points fed into the circuit in shown in Figure \ref{fig:innrprd} and ran one thousand twenty four times. This calculation only yields the square of the inner product. Since $|\braket{ E(D_i) }{ E(D_i) }| = 1$ for all data records $D_i$, to calculate the Euclidean distance Equation \ref{eq:inner-L2} is applied.

To examine how the geometry of the original data is altered with each encoding method, as well the difference of the geometry between each encoding methods, the Betti number of $b_0$ and $b_1$ are computed with increments of $.05$. Figure \ref{fig:compare} (a) displays the evolution of $b_0$, the number of connected components. Interestingly, there is a consistent difference in each Betti number, where around when $\epsilon = .5$ there is stability between the original data and the encoding methods. This consistency shows the encoding methods map the original data rigidly into the respective geometry, however, the information is fairly limited.

Figure \ref{fig:compare} (b) shows the Betti number density for the original data and the three encoding approaches. The $\epsilon$-thresholds start at $0.0$ and go to $1.0$. Unlike the Betti number for $b_0$, there are prominent differences between the original data and all of the encoding approaches. In particular, as the Betti number for the original data stabilize there is both an increase and decrease of the Betti number for the encoding methods, all of which eventually stabilize. This inconsistency displays how each technique effects the noise in the data in different ways.

The difference in the Betti numbers between the three techniques and original data are a bit surprising as one would expect the structure of the data to be intact since the each data point is mapped to a unitary operator, which keeps the underlying structure. However, since each data point is encoded into a unitary operator, it would be more applicable to compare each of these encoded data points as operators within this Lie group than to consider the output of each operator as a point in a Hilbert space.

While the Betti numbers show glaring discrepancies, it does not give insight into the respective geometries. Figure \ref{fig:barcodes} gives the barcode and diagrams for each encoding method and the original data. The barcodes display both subtle and stark contrasts of the geometries, putting more context to the counter-intuitive results of Betti numbers. The diagrams exhibit the 'expressibility' mentioned in  Schuld et al. \cite{schuld2021effect}, as the IQP method yields an intricate geometry. Interestingly, the amplitude method, while expressive given the intricacy of the control gates, yielded a simplistic geometry. The barcodes and diagrams were calculated using the Gudhi package \cite{gudhirips}.

\section{Discussion} 
The technique in this manuscripts displays how to apply qTDA in a hybrid manner, with the quantum advantage in comparing the encoded data points and calculating the eigenvalues of the corresponding Hamiltonian. The results of comparing the three encoding techniques and the data in calculating the Betti number for $b_0$ and $b_1$ showed discrepancies in information retention. It is posited that the encoded data must be considered as unitary operators in the Lie group $\mbox{SU}(2^n)$ and compared within the respective noncommutative geometry. The code used in Section \ref{sec:emp} is available on request. 

\section{Disclaimer}
About Deloitte: Deloitte refers to one or more of Deloitte Touche Tohmatsu Limited (“DTTL”), its global network of member firms, and their related entities (collectively, the “Deloitte organization”). DTTL (also referred to as “Deloitte Global”) and each of its member firms and related entities are legally separate and independent entities, which cannot obligate or bind each other in respect of third parties. DTTL and each DTTL member firm and related entity is liable only for its own acts and omissions, and not those of each other. DTTL does not provide services to clients. Please see www.deloitte.com/about to learn more.

Deloitte is a leading global provider of audit and assurance, consulting, financial advisory, risk advisory, tax and related services. Our global network of member firms and related entities in more than 150 countries and territories (collectively, the “Deloitte organization”) serves four out of five Fortune Global 500® companies. Learn how Deloitte’s
approximately 330,000 people make an impact that matters at www.deloitte.com. 
This communication contains general information only, and none of Deloitte Touche Tohmatsu Limited (“DTTL”), its global network of member firms or their related entities (collectively, the “Deloitte organization”) is, by means of this communication, rendering professional advice or services. Before making any decision or taking any action that
may affect your finances or your business, you should consult a qualified professional adviser. No representations, warranties or undertakings (express or implied) are given as to the accuracy or completeness of the information in this communication, and none of DTTL, its member firms, related entities, employees or agents shall be liable or
responsible for any loss or damage whatsoever arising directly or indirectly in connection with any person relying on this communication. 
Copyright © 2022. For information contact Deloitte Global.

\bibliographystyle{unsrt}
\bibliography{EncodingBib}
\end{document}